\documentclass{aa}
\usepackage{psfig}
\newcommand{\gauche}[1]{\begin{flushleft}\vspace{-4ex}#1\vspace{-2ex}\end{flushleft}}
\newcommand{\ascreta}[4]{$#1^{\mbox{h}}#2^{\mbox{m}}#3^{\mbox{s}}\kern-4pt.\kern1pt#4$}
\begin{document}

\thesaurus{06     %
              (11.03.4)
              } %

\title{BeppoSAX observation of the rich cluster of galaxies Abell~85}
\titlerunning{BeppoSAX observation of Abell~85}
\author{G.B. Lima Neto\inst{1,2} \and V. Pislar\inst{2,3} \and J. Bagchi\inst{4}}
\institute{%
Instituto Astron\^omico e Geof\'{\i}sico/USP, av. Miguel Stefano 4200, 
04301-904 S\~ao Paulo/S.P., Brazil \and
Institut d'Astrophysique de Paris, CNRS, 98bis Bd Arago, F-75014 Paris, France \and
Universit\'e du Havre, 25, rue Philippe Lebon, F-76600 Le Havre, France \and
Inter-University Centre for Astronomy and Astrophysics (IUCAA), Post Bag
4, Ganeshkhind, Pune 411007, India
}
\offprints{G.B. Lima Neto (gastao@iagusp.usp.br)}
\date{Accepted ???. Received ????; in original form ????}

\maketitle

\begin{abstract}
We report the observation of the Intra-Cluster Medium (ICM) of Abell~85 by the
X-ray satellite BeppoSAX. We have both analysed the spectrum obtained in the
central 8 arcmin circular region centred on the Very Steep Spectrum Radio
Source (VSSRS) and the spectra from a number of sub-regions. Analysis of the
spectra allowed us to independently obtain new estimates of the temperature,
metallicity and line-of-sight hydrogen density column, both globally
($T=6.6\pm0.3$~keV, $Z = 0.38\pm0.06Z_{\odot}$ and $N_{\rm H} =
5.5^{+0.9}_{-0.7} 10^{20}$cm$^{-2}$) and locally. These measures are in good
agreement with previous measures based on ROSAT and ASCA data. In the region of
the VSRSS, we have tried to disentangle the thermal from the non-thermal X-ray
emission. Although we could not do this unambiguously, we have nonetheless
estimated the extended magnetic field using the radio spectrum available for
this region. We obtain a lower limit intensity of $0.9 \mu$G,
consistent with our previous estimate. We also derive $\alpha$-elements/iron
abundance ratios that turn out to be higher than 1. Such a result tends to
support the burst model for elliptical galaxies, where a strong galactic wind
develops early in the galaxy history and type II supernovae (SN) may have the
main role in the enrichment of the ICM. A two-temperature ICM model was fitted
in the central region yielding a main component with roughly the mean cluster
temperature and a cooler component with temperature less than 0.1~keV.
\end{abstract}


\section{Introduction}

The hot ($T \approx 10^{8}$K) and tenuous (central density, $n_{0}\approx
10^{-3} $cm$^{-3}$) X-ray emitting gas found in rich clusters of galaxies is an
excellent tool to probe the cluster dynamics, morphology, and history. The
intra-cluster gas accounts for $\sim$ 10--15\% of the total cluster mass and
thanks to its short relaxation time scale, it can track the cluster global
gravitational potential. The main emission mechanism of the gas is the thermal
bremsstrahlung that is proportional to $\sim n^{2} T^{1/2}$ ($n$ is the
numerical density of electrons). Besides density and temperature, the observed
thermal bremsstrahlung emission may give estimations of the gas metallicity and
of the line-of-sight column density of hydrogen. The abundance of metals in the
intracluster gas is a sign of early enrichment by material processed in stars.
The determination of precise abundances of iron and $\alpha$-elements (nuclei
formed by the fusion of $\alpha$-particles in massive stars) can provide strong
constraints for the early evolution of galaxies and the galaxy-cluster medium
interaction.

In the last couple of years, the cluster Abell~85 (richness class 1, cD type,
optical redshift 0.056) has been extensively studied (cf. Pislar et al. 1997;
Lima Neto et al. 1997; Bagchi et al. 1998). Using data from the ROSAT satellite
detectors PSPC (imagery and spectroscopy) and HRI (high resolution imagery), it
was possible to draw substantial new conclusions about this cluster. Through the
use of wavelet techniques, Lima Neto et al. (1997) were able to detect a central
excess in the X-ray emission which has been identified with a cooling-flow of
about 50--150 M$_{\odot}/$yr (based on a multi-phase model of the gas).
Moreover, they have confirmed the presence of 3 small X-ray features around the
centre (cf. Prestwich et al. 1995) which may be a sign of inhomogeneities in the
central cooling-flow. The second most important X-ray feature in Abell~85 is a
blob to the south of the main structure (called the `South Blob' hereafter).
This blob is definitely a sub-structure in the X-ray map and coincides with a
small group of galaxies (the second brightest cluster member is there). However,
these galaxies do \textit{not} seem to form a gravitationally bound group and
may be part of a chain of galaxies extending to the neighbouring cluster
Abell~87 (Durret et al.~1998). The radio maps of the South Blob also show
intense, very steep spectrum radio emission at meter wavelengths (the
VLA map, see below and also Joshi et al.~1986; Bagchi et al.~1998). Most
of this emission is diffuse in nature and extends over 4 arcmin ($\sim 400\,$kpc)
but not associated with any particular galaxy. This, and another similar diffuse
radio source $\sim 1\,$Mpc to north-west (VSSRS 0038-096) are the principal
unexplained radio features in Abell 85.

The radio observations of diffuse synchrotron radiation strongly suggest the
presence of large scale magnetic field, $B$, and relativistic electrons in
clusters of galaxies. These electrons interact not only with magnetic fields but
also with the 3K cosmic microwave background radiation (CMB). The scatter of the
electrons and the CMB through inverse Compton effect (IC/3K) will produce X-ray
photons (e.g. Feenberg \& Primakoff 1948; Rephaeli \& Gruber 1988). The best
place to look for a co-spatial IC/3K and synchrotron radiation is on a very
steep spectrum radio source (VSSRS). These sources, possibly the remnants of
former radio galaxies but presently not identifiable with galaxies, are commonly
known as cluster `radio-relics' and `radio-haloes' (cf. Feretti \& Giovannini
1996 and Kronberg 1994 for reviews). They owe their steep radio spectra to
radiative energy losses but are prevented from rapid fading, due to expansion,
by the thermal pressure of the surrounding intra-cluster gas (Baldwin \& Scott
1973).
These VSSRS are also possibly the tracers of large scale shock waves that form
at the intersection of filaments and sheets of galaxies due to gravity driven
supersonic flows of extragalactic matter (En{\ss}lin et al. 1998; Miniati et al.
2000; Bagchi et al. 2000).

We have studied the X-ray excess at the location of the extended VSSRS 0038-096
in the western part of Abell~85 using ROSAT PSPC data and the low frequency
radio data (Bagchi et al.~1998). Even though ROSAT has a limited spectroscopic
range [0.5--2.4]~keV, we were able to derive a value for the magnetic field, $B
= 1.0 \pm 0.1\; \mu$G. We based our estimate on the assumption that the
co-spatial radio and X-ray emission are nonthermal synchrotron and IC/3K
radiation, respectively, from a common population of relativistic charges.

In this paper, we present new results based on observation of Abell~85 by the
X-ray satellite BeppoSAX. We use the spatially resolved spectroscopy
capabilities of the MECS and LECS detectors to determinate independently the
temperature, metallicity and line-of-sight hydrogen density column.
Furthermore, we try to separate the thermal from the non-thermal X-ray emission
in the regions of the VSSRS and the South Blob, and we give lower limit
estimates of the extended magnetic field.
In the central region, we determine the properties of the cooling-flow, showing
that the X-ray emission in that region is compatible with a metal rich,
2-temperature plasma. Moreover, when the signal to noise ratio is high enough we
estimate the individual abundances of iron, nickel and $\alpha$-elements. Unless
otherwise stated, we assume $H_{0} = 50$~km~s$^{-1}$Mpc$^{-1}$ and $q_{0} = 0.5$
(i.e., 1 arcmin corresponds to $97 h_{50}^{-1}$ kpc).

\section{The data}\label{sec:data}

Abell~85 was observed in July 1998 by the BeppoSAX satellite (Boella et al.
1997a) with two of the narrow field instruments: the low-energy concentrator
spectrometer (LECS, sensitive in the [0.1--10.0]~keV range; Parmar et al. 1997)
and two units of the medium-energy concentrator spectrometer (MECS 2 and 3,
sensitive in the [1.3--10.5]~keV range; Boella et al. 1997b).

The observation was not pointed at the centre of the cluster, at the position of
the central cD galaxy, but at the position of the VSSRS 0038-096. The net
exposure times were $92\,708$ and $40\,810$~s for the MECS and LECS,
respectively.

\subsection{Data reduction}

The data have been pre-processed (linearized and cleaned) using the SAXDAS
Rev.~1.1 package in the FTOOLS 3.5 environment at the BeppoSAX Science Data
Centre (SDC) and then retrieved through their archive.

We have used the merged data from both MECS units; they were merged using the
SAXDAS ``meevelin''  program by the SDC (see Fiore et al. 1999).

In order to avoid severe vignetting effects, we have extracted data for
scientific analysis only inside a circle of 8 arcmin around the pointing axis of
the LECS and MECS instruments (Cusumano \& Mineo 1998).

The background spectra have been extracted from sky fields devoid of detectable
sources and using the same regions on the detectors as the ones used for
extracting spectra of Abell~85 (see below the definition of these regions).

\subsection{Data Analysis}

The total field of view of the MECS units is $0.5^\circ$, but they are supported
by a beryllium support that blocks photons with $E < 5$ keV; this effect is 
\textit{not} taken into account in the effective area files (Fiore, private 
communication). Therefore, we restricted our analysis to the area inside the 
support, i.e., up to 8 arcmin from the detector axis.

Both instruments, LECS and MECS, have moderate spatial resolution, $3.7'$ at
0.28~keV and $1.2'$ at 6.4 keV (FWHM) respectively (Fiore et al. 1999). Hence,
we have defined sub-regions in the field of view in order to perform separate
spectral analysis. The X-ray data in these regions were extracted with XSELECT
1.4. Table \ref{tab:regions} resumes the characteristics of the regions we have
used. 

\begin{figure*}[htb]
  \psfig{figure=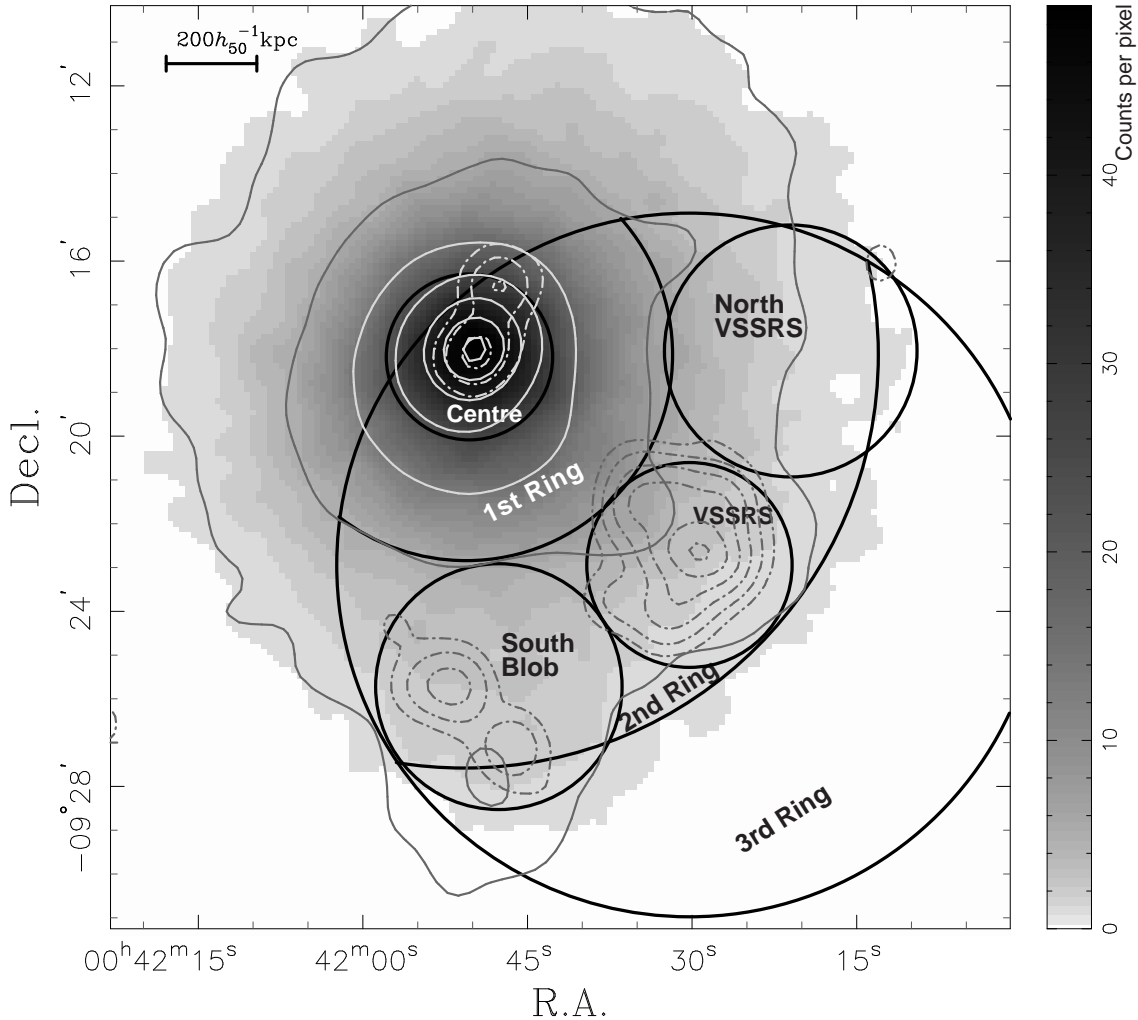,width=15cm} \caption[]{The regions used for
  the analysis of MECS and LECS data of Abell~85 are the thick black circles and
  rings. In grey-scale is the MECS23 image. The continuous thin lines (in white
  in the centre for the sake of visibility) are contours of the PSPC data
  smoothed with a $\sigma=0.45$ arcmin gaussian. The dot-dashed thin lines are
  contours from the VLA radio continuum map at 90~cm $\lambda$ (see text).}
  \label{fig:regionsBeppo}
\end{figure*}

Figure \ref{fig:regionsBeppo} shows the regions we have selected superposed on
the X-ray map derived from the MECS23 data. This figure also shows the 90~cm
radio continuum VLA and ROSAT PSPC maps (isocontour lines). For the radio data,
we have used the VLA C-configuration, archival interferometric data (programme
ID AR279). The data was obtained in July 1993 with total of 30 minutes of
integration on Abell~85 at the frequency of 333~MHz and bandwidth of 3.1~MHz.
For reduction, the NRAO `AIPS' software was employed. Following the usual
procedure of editing and amplitude and phase calibration, the calibrated data
was Fourier transformed and the $3 \times 3$ degree image was deconvoluted with
`CLEAN' algorithm. The final image was convoluted with a circular restoring beam
of 60 arcsec (FWHM) gaussian profile. Noteworthy are the strong radio emission
from the region of VSSRS, the South Blob and the cD galaxy at the cluster
centre. The PSPC X-ray map is fully described elsewhere (Pislar et al. 1997).

\begin{table*}[htb]
\caption[]{Regions used in the spatial/spectral analysis of Abell~85. The fluxes 
and luminosities are estimated using an absorbed MEKAL model. Errors are 
3$\sigma$.}

\begin{tabular}{l r r r p{7cm}}
\hline
Name & Count rate & Flux$^*$  & $L_{X}{}^\dagger$ & Localisation \\
     & (counts/s) & [2.0--10.0] keV & [2.0--10.0] keV\\
\hline

Centre & $0.259\pm0.002$ & $4.37\pm0.31$ & $6.13\pm0.41$ & \gauche{Circle at 
\ascreta{0}{41}{50}{27}, $-9^\circ18'12.0''$, radius $1.9'$.} \\

South Blob & $0.091\pm0.001$ & $1.53\pm0.21$ & $2.13\pm0.29$ & \gauche{Circle at 
\ascreta{0}{41}{44}{33}, $-9^\circ23'52.0''$, radius $3.2'$.} \\

VSSRS  & $0.0331\pm0.0006$ & $0.55\pm0.09$ & $0.77\pm0.12$ & \gauche{Circle at 
\ascreta{0}{41}{30}{27}, $-9^\circ22'56.0''$, radius $2.3'$.}\\

North VSSRS & $0.0362\pm0.0007$  & $0.61\pm0.11$ & $0.85\pm0.15$ & \gauche{Circle at
\ascreta{0}{41}{20}{95}, $-9^\circ18'04.0''$, radius $2.9'$.}\\

1st Ring  & $0.220\pm0.002$ & $3.70\pm0.27$ & $5.18\pm0.36$ & \gauche{Ring between $2.0'$ and $4.7'$, centred at
\ascreta{0}{41}{50}{27}, $-9^\circ 18'12.0''$. The region outside  $8'$ from the 
MECS axis is excluded.}\\

2nd Ring  & $0.111\pm0.001$ & $1.86\pm0.19$ & $2.60\pm0.25$ & \gauche{Ring between $4.7'$ and $9.4'$, centred at
\ascreta{0}{41}{50}{27}, $-9^\circ 18'12.0''$. The region outside  $8'$ from the 
MECS axis is excluded.} \\

3rd Ring  & $0.01176\pm0.0005$ & $1.98\pm0.21$ & $0.27\pm0.03$ & 
\gauche{Region outside the circle 
centred at  \ascreta{0}{41}{50}{27}, $-9^\circ 18'12.0''$ and radius $9.4'$,
excluding the region outside  $8'$ from the MECS axis.} \\
	
\hline
All Field & $0.587\pm0.003$ & $9.95\pm0.45$ & $13.94\pm0.61$ & \gauche{Circle at 
\ascreta{0}{41}{30}{27}, $-9^\circ 22'56.0''$ radius $8.0'$.} \\
\hline
\end{tabular}

$^*$ Flux is in units of $10^{-11}$erg~s$^{-1}$~cm$^{-2}$\\
$^\dagger$ Luminosity is in units of $10^{44}$ erg~s$^{-1}$
\label{tab:regions}
\end{table*}

We have extracted the spectra for each region and used simultaneously both the
LECS and MECS (units 2 and 3) data. Following the ``Cookbook'' (Fiore et al.
1999), we have used only the data in the interval [0.12--4.0] keV for the LECS
and [1.65--10.5] keV for the MECS.

For each region we have used a single temperature plasma, the absorbed
Raymond-Smith (Raymond \& Smith 1977) and the MEKAL (Kaastra \& Mewe 1993;
Liedahl et al. 1995) models. For some regions (centre, South Blob and VSSRS) we
have also modelled the intracluster medium with a combination of two models:
either two MEKAL or a MEKAL plus a power-law, this to take into account the
non-thermal emission. The absorption is due to the cold gas in the
line-of-sight, mainly hydrogen and helium. We have used the photoelectric
absorption cross-sections given by Balucinska-Church \& McCammon (1992).

When the counts were high enough, we also have tried the VMEKAL model, that is,
with a variable individual abundances rates for the elements that contributed
for the X-ray flux at the [0.1--10.0] keV band. Unfortunately, most of the
metals do not produce strong enough lines for unambiguous detection with
BeppoSAX. Therefore we have fixed the abundance values of He, C, N, O, Ne, Na,
Mg, and Al to 0.3$Z_{\odot}$.
 
The spectral fits were done using XSPEC v10.0. The BeppoSAX narrow field
instruments have channels of equal energy width, but a spectral resolution that
scales roughly with the square root of the energy (Boella 1997a). Therefore, if
a spectrum must be rebinned (in order to increase the counts per bin), one
should first group the bins taking into account the energy resolution
non-linearity rather than simply grouping the channels with some \textit{ad hoc}
prescription. We have used the appropriate rebinning template files made
available by the SDC (Fiore et al. 1999).

However, even with the energy-dependent rebinning, there are still bins with low
counts. We have then used the recipe given by Churazov et al. (1996) based of
the smoothed rebinned spectrum (available in XSPEC) for computing the
statistical weights. With this procedure, one can still use least-square
minimisation fit and $\chi^{2}$ statistics for unbiased parameter and error
estimation.

\section{Results}\label{sec:results}

\begin{table*}[htb]
\tabcolsep=0.56\tabcolsep
\caption{Spectral fitting results for single component 
models. Errors are $3\sigma$ except when explicitly stated. The abundances are 
given in Solar units using values from Anders \& Grevesse (1989).
All metals, other than those in the table, are fixed at $Z=0.3 Z_{\odot}$.}
\begin{tabular}{r r c c c c c c c c c c c}
\hline
Region & Model & $N_{\rm H}$ & $kT$ & $Z$ & $Z_{\rm Si}$ & $Z_{\rm
S}$ & $Z_{\rm Ar}$ & 
$Z_{\rm Ca}$ & $Z_{\rm Fe}$ & $Z_{\rm Ni}$ & $\chi^{2\strut}$/dof \\
  &  &  ($10^{20}$cm$^{-2}$) & (keV) \\
\hline

all field & \textsc{mekal}  & $5.5^{+0.9}_{-0.7}$ & $6.6^{+0.3}_{-0.3}$ & 
 $0.38^{+0.06}_{-0.06}$ &---&---&--- &--- & ---&---& {\small 205.1/178} \\[2pt]

all field & \textsc{vmekal} & $5.5^{+1.0}_{-0.8}$ & $6.7^{+0.4}_{-0.3}$ &  & 
$0.6^{+0.3}_{-0.3}{}^*$ & $0.2^{+0.3}_{-0.2}{}^*$ & $0.3^{+0.6}_{-0.3}{}^*$ &
$0.6^{+0.6}_{-0.6}{}^*$ & $0.30^{+0.05}_{-0.05}$ & $0.1^{+0.4}_{-0.1}{}^*$ & 
{\small 202.7/173}  \\[2pt]

 Centre & \textsc{mekal} & $9.5^{+3.8}_{-2.7}$ & $6.2^{+0.5}_{-0.4}$ & 
 $0.48^{+0.10}_{-0.09}$ &  &  &  &  & & & {\small 198.9/161} \\[2pt]

 Centre & \textsc{vmekal} & $8.7^{+4.2}_{-2.5}$ & $6.2^{+0.6}_{-0.5}$ &  &
 $0.7^{+0.4}_{-0.4}{}^*$ & $0.6^{+0.5}_{-0.4}{}^*$ & $1.0^{+0.9}_{-0.8}{}^*$
 & $2.4^{+0.9}_{-0.8}{}^*$ & $0.40^{+0.09}_{-0.08}$ & $0.4^{+0.6}_{-0.4}{}^*$ &
 {\small 192.2/156}\\[2pt]

 1st Ring & \textsc{mekal} & $5.1^{+1.2}_{-0.9}$ & $6.6^{+0.6}_{-0.5}$ & 
 $0.34^{+0.10}_{-0.09}$ &---&---&---&---& ---& ---& {\small 214.4/178} \\[2pt]

 1st Ring & \textsc{vmekal} & $5.0^{+1.2}_{-0.9}$ & $6.7^{+0.6}_{-0.6}$ &  &
 $0.6^{+0.4}_{-0.4}{}^*$ & $0.3^\dagger$ & $1.0^{+0.9}_{-0.9}{}^*$ & 
 $0.3^\dagger$ & $0.28^{+0.08}_{-0.08}$ & $0.1^{+0.6}_{-0.1}{}^*$  & 
 {\small 212.72/175}\\[2pt]

 2nd Ring & \textsc{mekal} & $3.6^{+1.5}_{-1.0}$ & $6.7^{+0.9}_{-0.8}$ & 
 $0.20^{+0.13}_{-0.13}$ &--- &---&---&---&---&---& {\small 120.4/142} \\[2pt]

 2nd Ring & \textsc{vmekal} & $3.5^{+1.6}_{-1.0}$ & $6.7^{+1.1}_{-0.9}$ &  &
 $0.8^{+0.6}_{-0.6}{}^*$ & $0.5^{+0.6}_{-0.5}{}^*$ &$0.3^\dagger$ & 
 $0.6^{+1.3}_{-0.6}{}^*$ & $0.17^{+0.11}_{-0.11}$ & $0.8^{+0.9}_{-0.8}{}^*$ &
 {\small 118.90/138} \\[2pt]

 3rd Ring & \textsc{mekal} & $5.5^{+4.8}_{-2.4}$ & $9.1^{+6.9}_{-2.8}$ & 
 $0.6^{+0.6}_{-0.6}$ &---&---&---&---&--- &--- & {\small 79.6/72} \\[2pt]
 
 S. Blob & \textsc{mekal} & $4.6^{+1.9}_{-1.2}$ & $6.9^{+1.0}_{-0.8}$ & 
 $0.24^{+0.15}_{-0.14}$  &--- &--- &---&---&---&---& {\small 127.7/108} \\[2pt]

 S. Blob & \textsc{vmekal} & $4.5^{+1.9}_{-1.2}$ & $6.8^{+1.2}_{-0.9}$ & 
  & $0.1^{+0.6}_{-0.1}{}^*$ & $0.5^{+0.7}_{-0.5}{}^*$ & $0.3^\dagger$
  & $1.7^{+1.6}_{-1.3}{}^*$ & $0.20^{+0.12}_{-0.11}$  & $0.6^{+1.0}_{-0.6}{}^*$ 
  & {\small 126.03/104} \\[2pt]

 VSSRS & \textsc{mekal} & $3.3^{+2.8}_{-1.4}$ & $6.6^{+1.8}_{-1.3}$ & 
 $0.31^{+0.27}_{-0.25}$  &---&---&---&---&--- &---& {\small 104.8/108} \\[2pt]
 
 N. VSSRS & \textsc{mekal} & $2.6^{+2.9}_{-1.3}$ & $7.1^{+2.1}_{-1.4}$ & 
 $0.33^{+0.28}_{-0.27}$  & ---&--- & ---&---&---& ---& {\small 117.8/108} \\[2pt]
 \hline
\end{tabular}

${}^*$  $1 \sigma$ error\\
${}^\dagger$ Value fixed.
\label{tbl:fitsResults}
\end{table*}

\subsection{All field}

Within the unobstructed field-of-view of a circle of 8 arcmin centred at the
MECS axis we have our greatest signal to noise ratio and in this region we can
obtain some well constrained mean quantities for Abell~85. Using a MEKAL model,
we find a temperature of $6.6\pm0.3$~keV, an hydrogen column density
$N_{H}=5.5^{+0.9}_{-0.7}10^{20}$cm$^{-2}$ and a metallicity of $0.38\pm0.06
Z_{\odot}$ (cf. Table~\ref{tbl:fitsResults}). We obtain a significantly higher
temperature than that obtained with the ROSAT PSPC ($4\pm1$ keV, Pislar et al.
1997), but essentially the same as the one obtained with ASCA ($6.1\pm0.2$~keV,
but notice that this is the mean temperature within $\sim 15$ arcmin; Markevich
et al. 1998). The column density is higher than the Galactic value deduced from
HI data in the field of view (for example $3.08 \times 10^{20}$cm$^{-2}$ at the
position of the VSSRS and $3.58 \times 10^{20}$cm$^{-2}$ at the position of the
cD, Dickey \& Lockman 1990). This suggests that there is an amount of HI
contained within the cluster itself.

The metallicity is well constrained and has the value generally found in this
type of clusters (e.g. Fukazawa et al. 1998). We have further fitted the data
with a VMEKAL model to obtain abundances of the individual metal elements. The
reduced $\chi^2$ ($\chi^2$ divided by the number of degrees of freedom) is
slightly higher for the VMEKAL compared to the MEKAL model, but the temperature
and hydrogen column density are the same and we detect the presence of some
metals (cf. Table~\ref{tbl:fitsResults}). Only the abundances of Ni and Fe are
not compatible with zero at 1$\sigma$ level and, except for these two elements,
we can only give upper limits for the abundance.

The total X-ray luminosity in the [0.1--2.4] keV band is $(8.49\pm0.37)\
10^{44}h^{-2}_{50}$ erg~s$^{-1}$ inside a radius of $770h^{-1}_{50}$kpc (8
arcmin) centred at the VSSRS position.

\subsection{Centre}

\begin{figure}[htb]
\psfig{figure=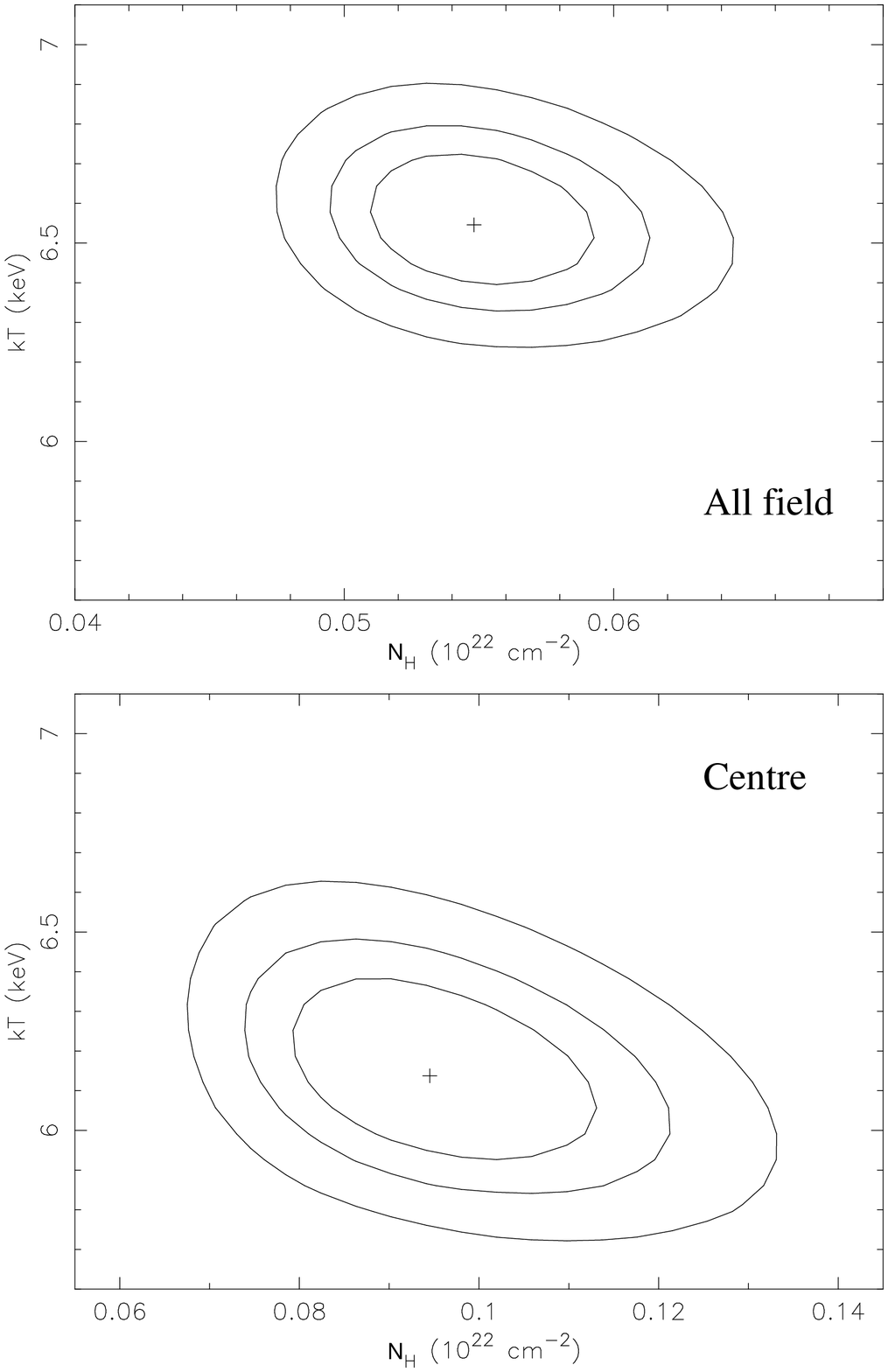,width=7.5cm}
\caption{Contours of chi-square values for the $N_{\rm H}$--$kT$ plane. These
contours correspond to the best fits  with the \textsc{mekal} model (cf.
Table~\ref{tbl:fitsResults}.) for ``all field'' (upper panel) and the centre of the
cluster (lower panel). Notice that the $n_{\rm H}$ scale is different for the
two plots. The contours shown are at the $1\sigma$, $2\sigma$, and $3\sigma$ confidence
level.}
\label{fig:nH_T_all_centre}
\end{figure}

We take a region of 1.9 arcmin ($185 h_{50}^{-1}$ kpc) at the position of
the cD galaxy located at the cluster centre. A fit with a single temperature
model shows a significant increase of the hydrogen column density and a decrease
of temperature relative to the values obtained when we fitted the whole cluster.
In a previous study, Pislar et al. (1997) observed a negative correlation
between the X-ray gas temperature and $N_{\rm H}$. Here, we observe a much weaker
correlation (Fig.~\ref{fig:nH_T_all_centre}), so the effect -- temperature
decrease inwards the centre -- is probably real and is not an artefact of the
fit. The metallicity is significantly higher in the central region than in the
whole field. The ROSAT data gave the same results (Pislar et al. 1997). We
detect the presence of Si, S, Ar, Ca and Fe when we fit the data with a VMEKAL
model. These results can be interpreted as due to the accumulation of neutral
hydrogen in the vicinity of cD galaxy, due to the presence of cooling flow with
star formation.

\begin{figure*}[htb]
	\mbox{\psfig{figure=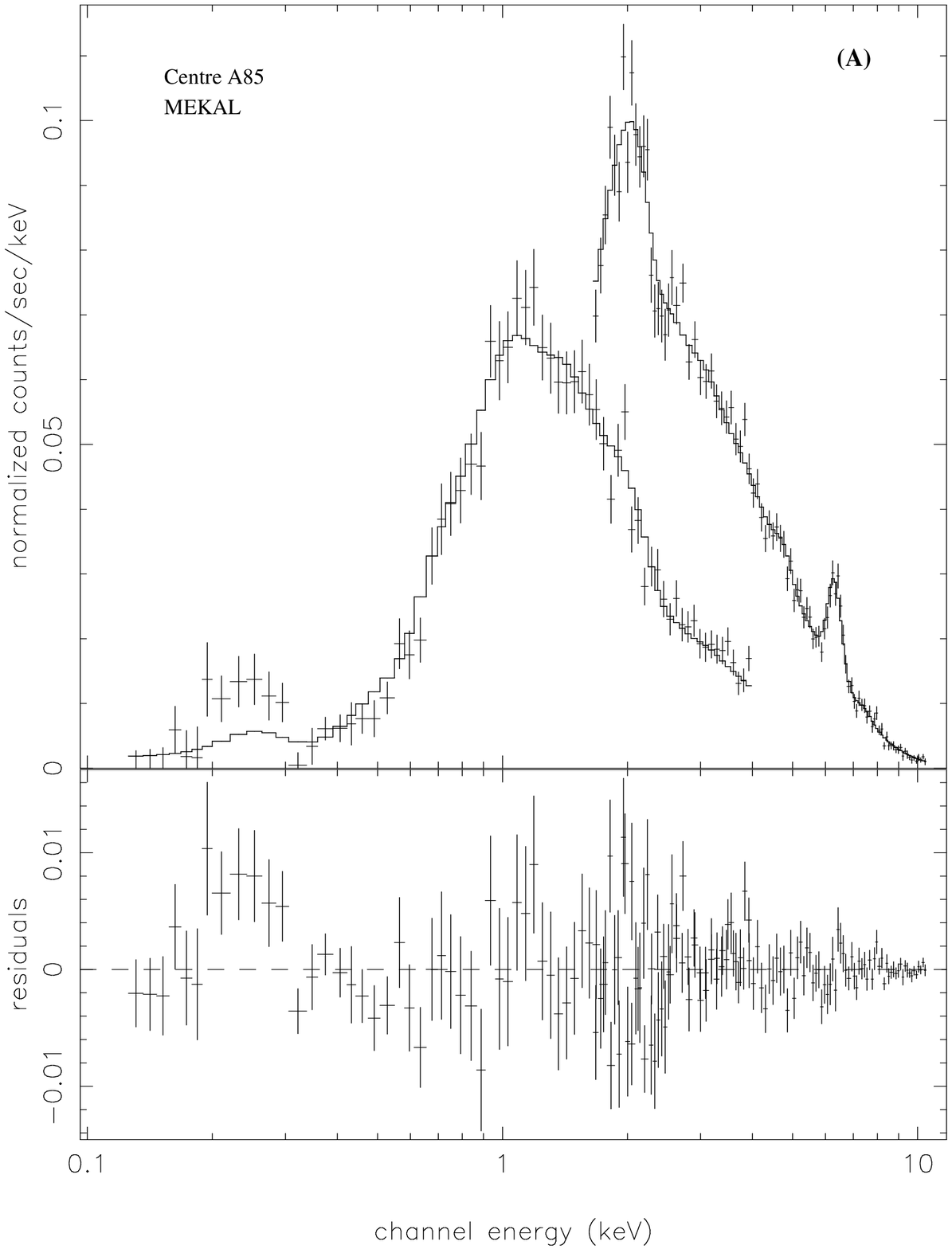,width=8cm}\hspace{16pt}
	\psfig{figure=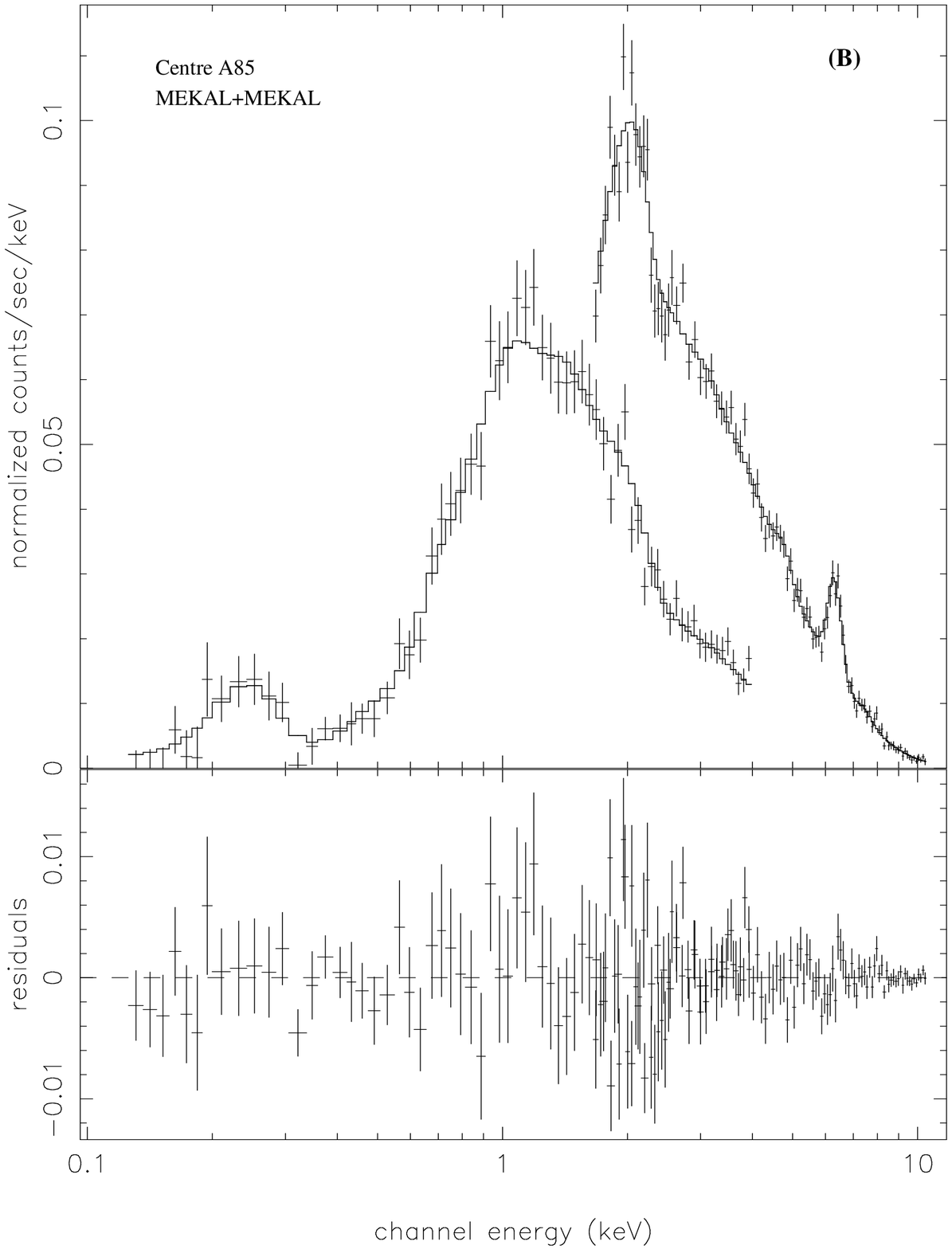,width=8cm}}
	\caption{Plasma model fits to the centre region of Abell~85. Left:
        Single temperature MEKAL model. Right: two temperature
        model.
	Notice the low energy photon excess in panel (A) compared to the two
	temperature model (B).}
	\label{fig:spectreCentre}
\end{figure*}

However, a single temperature plasma model is unable to provide a good fit for
the spectrum of the central region of Abell~85 when the photon energy is less
than 0.4~keV: as is shown in figure \ref{fig:spectreCentre}(A), there is a clear
excess of soft X-ray photons.

Indeed, previous studies of Abell~85 have suggested that this cluster houses a
strong central cooling flow (e.g. Stewart et al. 1984; Edge et al. 1992;
Prestwich et al. 1995; Pislar et al. 1997; Lima Neto et al. 1997). Estimated
values range from about 50 to 200 $M_{\odot}$/yr inside a cooling radius of
$\sim 100$--$200 h_{50}^{-1}$kpc. We can thus interpret the soft X-ray excess
component as evidence for the presence of cool gas. Therefore we have fitted the
central region spectrum with a 2-temperature gas model.

When we fix 2 or 3 parameters of the first thermal component, using the ``all
field'' mean values, the fits are not significantly improved (cf.
Table~\ref{tbl:2CompCentre}). But when we let all the parameters free to vary,
the fit is significantly improved (the reduced $\chi^{2}$ is 1.13 instead of
1.23). Figure \ref{fig:spectreCentre}(B) shows the best 2-temperature gas model
fit. The parameters of the first component are roughly the same as that obtained
in a single component fit, but the temperature of the second component is found
to be very small (less than 0.1~keV at $3\sigma$ level). Thus, we recover the
previous results based on the PSPC data reported in Lima Neto et al. (1997), in
particular their 2-temperature fit of the central region, with the cooler
component at $T=0.10\pm0.03$~keV (see Table~2 in their paper).

Since the central cD galaxy is also a radio source (as can be seen in
Fig.~\ref{fig:regionsBeppo}) we have tried a fit with a thermal component and a
power law, in order to model an eventual non-thermal X-ray source (e.g. an
Inverse Compton scattering of relativistic electrons with the microwave
background photons). However, the inclusion of a power-law component does not
improve the fit and we cannot detect such an emission.

\begin{table*}[htb]
\tabcolsep=0.56\tabcolsep
\caption[]{Two-component gas model results. The flux is without absorption 
in units of $10^{-11}$erg~s$^{-1}$~cm$^{-2}$, where Flux$_{1}$ and Flux$_{2}$
refer either to the thermal and non-thermal components, respectively, or
to two thermal components (in the \textsc{mekal+mekal} case).
`\textsc{2pow}' means broken power-law. Errors are $3\sigma$ except when explicitly stated.}
\begin{tabular}{l l c c c c c c c c c}
\hline
Region & Models &
$N_{H}$ & $T_{1}$ & $Z_{1}$ & $T_{2}$ & $Z_{2}$ & $\alpha$ & 
 Flux$_{1}$ & Flux$_{2}$ & $\chi^2/$dof \\
 & & ($10^{20}$cm$^{-2}$) & (keV)  &($Z_{\odot}$) & (keV) & ($Z_{\odot}$) & 
& [2--10] keV  &[2--10] keV  \\
\hline

Centre & \textsc{mekal+mekal} & $5.5^\dagger$ & $6.6^\dagger$ & 
$0.38^\dagger$ & $5.9_{-2.0}^{+3.7}$ & $1.0^\dagger$ &---& $3.55\pm0.25$ & $0.86\pm0.15$
& 228.41/162\\[2pt]

Centre & \textsc{mekal+mekal} & $9.5_{-2.7}^{+3.8}$ & $6.6^\dagger$ & 
$0.38^\dagger$ & $6.1_{-3.8}^{+5.5}$ & $>0.4$ &---& $0.52\pm0.04$ &
$3.89\pm1.00$ & 198.98/160 \\[2pt]

Centre & \textsc{mekal+mekal} & $11.8_{-2.9}^{+3.4}$ & $6.0_{-0.4}^{+0.4}$ & 
$0.48_{-0.09}^{+0.10}$ & $2.2_{-1.3}^{+6.7} 10^{-2}$ & $>0.0$ &---& $4.42\pm0.14$ &
$<0.01$ & 178.70/158 \\[2pt]

Centre & \textsc{mekal+pow} & $9.4_{-2.6}^{+4.5}$ & $6.2_{-0.4}^{+0.5}$ & 
$0.48_{-0.09}^{+0.10}$ & ---&---& $1.0^\dagger$ & $4.42\pm0.14$ & $<0.01$ & 198.93/160 \\[6pt]

S. Blob  & \textsc{mekal+pow} & $5.3_{-1.9}^{+8.1}$ &$7.1_{-3.1}^{+1.7}$ & 
$0.25_{-0.16}^{+0.20}$ &---& ---& $1.3^{+1.7}_{-1.3}{}^*$  & $1.47\pm0.24$ &
$0.064\pm0.057 {}^*$ & 127.05/106 \\[6pt]

VSSRS & \textsc{mekal+pow} & $3.8_{-1.8}^{+5.2}$ & 
$6.7_{-1.4}^{+2.8}$ & $0.33_{-0.27}^{+0.59}$ &---&---& $1.0^\dagger$ &
$0.52\pm0.09$ & $0.033{}^\dagger$ & 104.37/107 \\[2pt]

VSSRS & \textsc{mekal+pow} & $3.5_{-1.5}^{+5.7}$ & 
$5.5_{-2.6}^{+4.3}$ & $0.35_{-0.29}^{+1.59}$ &---&---& $0.4_{-0.4}^{+4.8}$ &
$0.44\pm0.14$ & $0.12\pm0.12$ & 104.04/106 \\[2pt]
\hline

 &  &
$N_{H}$ & $T_{1}$ & $Z_{1}$ & $\alpha_{1}$ & Break & $\alpha_{2}$ & 
 Flux$_{1}$ & Flux$_{2}$ & $\chi^2/$dof \\
 & & ($10^{20}$cm$^{-2}$) & (keV)  &($Z_{\odot}$) &   & (keV) & 
& [2--10] keV  &[2--10] keV  \\
\hline

VSSRS & \textsc{mekal+2pow} & $4.0_{-1.5}^{+1.9}$ & $7.3_{-1.4}^{+2.7}$ & 
$0.34_{-0.17}^{+0.23}$ & 1.0${}^\dagger$ & $>0$ & 1.85${}^\dagger$ & 
$0.52\pm0.09$ & $0.03\pm0.03$ & 103.71/106\\[2pt]

VSSRS & \textsc{mekal+2pow} & $3.7_{-1.0}^{+1.5}$ & $3.9_{-1.4}^{+5.7}$ &
$0.44_{-0.25}^{+0.43}$ & 0.4${}^\dagger$ & $7.4_{-0.7}^{+0.9}{}^*$ &
1.85${}^\dagger$ & $0.27\pm$0.18 & $0.28\pm0.01{}^*$& 101.22/106\\

\hline
\end{tabular}

${}^*$  $1 \sigma$ error\\
${}^\dagger$ Value fixed\\
\label{tbl:2CompCentre}
\end{table*}

\subsection{South Blob}

The South Blob region is a circle of 3.2 arcmin centred at the coordinates
indicated in Table~\ref{tab:regions}. We have fitted this region with both MEKAL
and VMEKAL single temperature models (Table~\ref{tbl:fitsResults}) and a MEKAL
plasma superposed to a power-law emission (Table~\ref{tbl:2CompCentre}). The
later was motivated by the presence of an extended radio emission seen on the
90cm VLA map (Fig.~\ref{fig:regionsBeppo}) and on the 326.5 MHz OSRT map (Fig.~1
in Bagchi et al. 1998). Such an extended radio emission reveals the presence of
relativistic electrons that will interact with the microwave background photons
and produce an Inverse Compton X-ray emission, depending on the (a priori
unknown) Lorentz factor and energy distribution of these electrons.

With both single temperature models, the hydrogen column density is higher than
in previous fits (Pislar et al. 1997) and marginally compatible with the
Galactic value ($3.08\times 10^{20}$cm$^{-2}$, Dickey \& Lockman 1990). The
hydrogen column density is even higher, however, when we use a MEKAL plus
power-law model (although still compatible within their error bars). The
temperature is higher than in the centre, being actually above the cluster mean
temperature. With the MEKAL plus power-law model, the obtained temperature is
even higher (but with larger error bars). Notice that, comparing the reduced
$\chi^2$, an addition of a power-law component does not improve the fit.

The metallicity obtained in the South Blob is significantly lower than the mean
value obtained for the whole cluster and is about half the value found in the
centre of the cluster. This lower metallicity is also obtained with the iron
abundance in the VMEKAL model. Besides the iron, only Ca was detected with an
abundance $1\sigma$ above zero (but with a very large upper limit).

\subsection{VSSRS}

This is the region where a diffuse, very steep spectrum radio source (MRC
0038-096; see Bagchi et al. 1998) is observed.

Here, fitting a single plasma component, we obtain the same temperature and
metallicity as in the whole field (cf. Table \ref{tbl:fitsResults}), even though
the error bars are significantly larger due to the smaller count rate. The
hydrogen column density is the same as the Galactic value.
 
In both the HRI and PSPC images it is possible to notice an X-ray excess in this
region, spatially correlated with the diffuse VSSRS (Lima Neto et al. 1997).
This X-ray excess, interpreted as IC/3K emission, was estimated by Bagchi et al.
(1998), by subtracting the total X-ray emission by the thermal contribution. The
latter was computed using a detailed plasma model for Abell 85 (see Pislar et
al. 1997).

\begin{figure}[htb]
 \psfig{figure=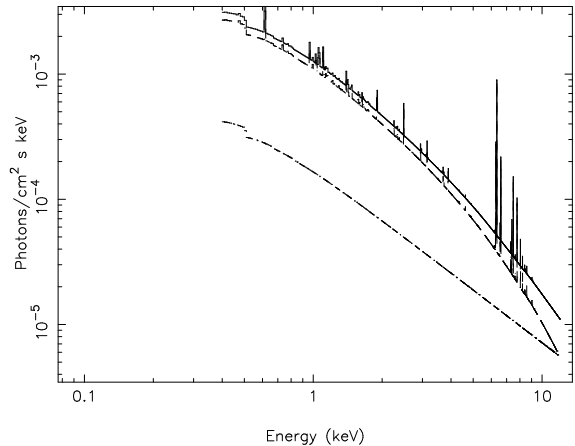,width=7.5cm}
 \caption{Plasma model with two components, MEKAL and power-law, for the
 VSSRS region. The lower dash-dotted curve represents an Inverse Compton
 emission with spectral index $\alpha=0.4$; the dashed curve is a MEKAL model
 with temperature 5.5 keV and abundance 0.35 $Z_\odot$ (cf.
 Table~\ref{tbl:2CompCentre}). The continuous line is the sum of the
 components.}
\label{fig:vssrsModelMekal}
\end{figure}
 
Ideally, one would like to disentangle directly and unambiguously, the IC/3K
from the thermal X-ray emission. In principle, this could be done by fitting a
composite model to the X-ray spectrum. Thus, we have fitted the X-ray spectrum
in the VSSRS region with the sum of a MEKAL plus a power-law model, the latter
representing the IC/3K emission. The results are summarized in Table
\ref{tbl:2CompCentre}. We have obtained, for the IC/3K emission, a spectral
index of 0.4 but the reduced $\chi^2$ for the composite model is not
statistically better than the one of a single MEKAL model. In
Fig.~\ref{fig:vssrsModelMekal} we show the best fit model for the VSSRS.
Extrapolating the model to higher energies, the IC/3K mechanism starts to
dominate the total X-ray flux beyond $\sim 15 $keV (a well know result), outside
the range of the MECS detector.

The difficulty found in disentangling the thermal from the non-thermal X-ray
emission is due to two main factors: (I) the IC/3K emission is relatively weak
compared to the thermal one and (II) the bremsstrahlung emission is well
approximated by a power-law in the energy interval observed by BeppoSAX.

We have also tried a fit with a fixed value for the spectral index equal to 1,
i.e., the same spectral index of the VSSRS between 10 and 100~MHz (see Fig.~3 of
Bagchi et al. 1998). Again, the fit is not significantly improved but the flux
of the IC/3K is not negligible. Notice, however, that the best fit spectral
index of 0.4 is consistent with the radio spectrum in the interval [10--40]~MHz,
where the spectrum seems to flatten.

Furthermore, we have modelled the IC/3K emission as a broken power-law,
characterized by two power-law slopes and a breaking point. We could not obtain
a robust fit letting the two slopes being free parameters. Therefore, we
fixed the steeper slope, $\alpha_{2}$, equal to the observed slope of the radio
spectrum in the range [100--400] MHz, i.e., $\alpha_{2}=1.85$ (Bagchi et al.
1998) and fixed the shallower slope ($\alpha_{1}$) either to 1.0 (from the radio
spectrum) or to 0.4, our best fit with a single power-law (cf.
Table~\ref{tbl:2CompCentre}).

In the first case, with $\alpha_{1}=1$, the fit is only slightly better than
those with a single power-law and the derived IC/3K flux is much smaller than
the thermal flux. Unfortunately, we cannot obtain a meaningful value for the
breaking point. In the second case, $\alpha_{1}=0.4$ the fit is still better and
the breaking point is well determined at 7.4~keV. However, the derived fluxes
for the thermal and the broken power-law components are almost the same,
indicating the confusion in separating the thermal from the non-thermal
emission.

Sarazin (1999) has computed the Inverse Compton spectra for a number of models
with different relativistic electron population. For the models with steady
particle injection, the IC emission spectra in the combined MECS and LECS range
would be well represented by a single power-law with slope $\alpha \approx
(p/2)$, where $p$ is the logarithm slope of the relativistic electron
distribution (cf. Fig.~12 of Sarazin 1999).

For the models with no injection of electrons but with an initial population,
the IC emission could only be detectable if the initial population were injected
in the cluster at redshift lower than 0.1 (cf. Fig.~13 of Sarazin 1999). 
In this case, a broken power-law is needed to reproduce the IC emission with 
$\alpha_{1} \approx (p-1)/2$ and $\alpha_{2} \approx p$ (actually the emission 
has an exponential cut-off and these slopes are only valid in the range 
[0.1--10.0] keV).

\subsection{Other Regions}

We have made fits in the regions we called 1st, 2nd, 3rd Rings and N. VSSRS. In
the 3rd ring and N. VSSRS regions, the error bars are quite large due to the
small number of counts and it is difficult to deduce strong results.

The hydrogen column density and metallicity decrease from the centre to the 2nd
ring, while the temperatures of the 1st and 2nd rings are quite larger than in the
centre.

\section{Discussion}

\subsection{Estimation of Magnetic Fields}\label{magnetic}

In section \ref{sec:results}, we have attempted to separate the thermal and
non-thermal X-ray emission components at the VSSRS and at the South Blob
regions, where radio observations have detected diffuse synchrotron radiation,
showing the presence of magnetic fields and relativistic electrons. Here, we
estimate the volume averaged $B$ field in these zones, using the non-thermal
fluxes (cf. Table~\ref{tbl:2CompCentre}). We wish to inject a
cautionary note at this stage: in this section the $B$ values are expressed as
actual estimates, where we have used the nominal results given in
Table~\ref{tbl:2CompCentre}. However, given the non-thermal X-ray fluxes
errors bars, these fluxes should be regarded as upper limits. Therefore all
our estimates of the $B$ field intensities should be ideally treated as only
lower limits.

\subsubsection{The VSSRS region}

In the first case (MEKAL+POW), when the spectral indices of non-thermal X-ray
and radio photons are both fixed at $\alpha_{1} = 1$, the IC/3K flux ($f_{\rm
IC}$) is estimated to be $3.3 \times 10^{-13}$erg~s$^{-1}$cm$^{-2}$ in 2--10~keV
range and the radio flux ($f_{\rm S}$) to be $6.6 \times
10^{-14}$erg~s$^{-1}$cm$^{-2}$ in 10--100~MHz range. This leads to a field value
of $B = 1.34\pm 0.24 \mu$G.

In the second case, when we have allowed the spectral index to vary, we have
obtained the value of $\alpha_{1} = 0.4$, with large error bars
(Table~\ref{tbl:2CompCentre}). This value of spectral index seems roughly
consistent with radio spectral shape in the range 10--45~MHz (Bagchi, Pislar \&
Lima Neto 1998). In this case we have $f_{\rm IC} = 1.2 \times
10^{-12}$erg~s$^{-1}$cm$^{-2}$ and $f_{\rm S} = 4.4 \times
10^{-14}$erg~s$^{-1}$cm$^{-2}$. From these, we then obtain $B = 0.40 \pm 0.13 \mu$G.

Finally, we consider the more complex model where the non-thermal spectrum is
represented by two power-law forms with a `spectral break' in-between
(MEKAL+2POW). When the low and the high frequency spectral segments are fixed
\textit{a priori}, with spectral indices $\alpha_{1} = 1$ and $\alpha_{2} = 1.85$, we
once more obtain the magnetic field as $B = 1.34 \pm 0.24 \mu$G (but note that the
`break' value for X-ray photons is not obtained).

However, when $\alpha_{1}$ is fixed at 0.4 and $\alpha_{2}$ is fixed at 1.85,
we obtain a well constrained value (7.4 keV) for the `break' in IC/3K spectrum
(Table~\ref{tbl:2CompCentre}). It is possible to relate this break to a similar
break in the radio spectrum since both come from a common electron energy spectrum. The
frequencies of IC/3K photons ($\nu_{IC}$) and of synchrotron photons
($\nu_{S}$) are related by:
$$\nu_{\rm IC}(\mbox{Hz}) = 4.89 \times 10^{10} \, (1+z) \, \nu_{\rm S}(\mbox{Hz})
\, [B(\mu \mbox{G})]^{-1} \, .$$
The decametric radio spectrum for the VSSRS shown in Figure 3 of Bagchi et al.
(1998) does show a sudden flattening or `break' in the spectrum at $\approx
30$--40~MHz. Assuming $\nu_{\rm IC} = 1.77 \times 10 ^{18}$~Hz (or 7.4~keV) and
$\nu_{\rm S} = (35 \pm 10)\times 10^{6}$~Hz, we have another magnetic field 
estimate for
the VSSRS from the last equation. This is obtained as $B = 1.01\pm 0.31 \mu$G.

Putting together the above results obtained, using the above mean values and
standard deviations, results in $B \ge 0.9 \mu$G. As per the caveat above,
this really is the lower limit to the magnetic field value and should be
treated as such. Our new estimate using the \textit{BeppoSAX} data is
consistent with our earlier estimate ($B = 1.0 \pm 0.1\; \mu$G) using ROSAT
data (Bagchi et al. 1998).

\subsubsection{The `South Blob' region}

Currently, due to insufficient radio data, the spectral index value for the
emission from the South Blob is not available. However, the radio data from Ooty
Synthesis Telescope at 327 MHz detects the total 450 mJy diffuse radio flux from
this zone (Bagchi et al. 1998). In section \ref{sec:data}, by model fitting, we
have obtained an estimate of $6.4 \times 10^{-13}$erg~s$^{-1}$cm$^{-2}$ for the
IC/3K flux and $\alpha = 1.3$ for the spectral index. Assuming this value for
the spectral index over the entire radio range (10 MHz to 10 GHz), we can then
easily calculate the lower limit value for magnetic field in this
region, which is $B \ge 0.4\,\mu$G.


\subsection{Cluster merging at the South Blob}

The temperature detected in the South Blob is higher than in the other regions
of Abell~85. This result is similar to the one found by Markevitch et al. (1998)
(cf. their Fig.~2) using ASCA data. They interpret the higher temperature in the
region of the South Blob with the model proposed by Durret et al.~(1998), that
is, of a substructure falling in the main body of Abell~85. The higher
temperature in this region, compared with either a region symmetrically opposed
towards the north of Abell~85 or a region farther to the south, may be explained
by a shock that heats the ICM. Although, with BeppoSAX, we cannot measure
temperatures southern from the South Blob or much northern from the centre, the
higher temperature that we do detect at the South Blob and the 2nd Ring tends to
confirm this picture of merging substructure.

We have also detected, with a $3\sigma$ level, an excess on neutral hydrogen
compared to the galactic value (from Dickey \& Lockman 1990). This excess, about
1.5 times the galactic value, also supports the scenario where gas is pouring in
this region. Notice, that the $N_{\rm H}$ measured over the entire 2nd Ring
(which contains most of the South Blob) is substantially closer to the galactic
value, suggesting that the localised excess at the South Blob region is real.

Interestingly, the South Blob presents a significantly lower abundance of metals
compared to the other regions of Abell~85 and to its mean value. The 2nd ring,
which contains the South Blob, has also the same low metal abundance. If we
admit the scenario of Durret et al. (1998), where a filament of very low surface
brightness is falling into Abell~85, along the axis coming from Abell~87 and,
moreover, Abell~87 is in fact composed by poor sub-clusters, then one can
suppose that the gas in the filament falling into Abell~85 has a rather low
metallicity. This could explain the low abundance determined at the South Blob
as the mixture of the $Z \sim 0.4 Z_{\odot}$ ICM from Abell~85 with a lower
metal rich gas falling along the filament.

If the South Blob is indeed a place where a substructure is merging
with the main body of Abell~85, then we may have a strong shock in this region.
Such a shock is bound to accelerate the ICM electrons to relativistic energies
and also may amplify stochastically the pre-existing magnetic fields (En{\ss}lin
et al. 1998; Sarazin 1999). These would in turn produce X-ray emission by IC/3K
scattering and the radio synchrotron radiation.
Our best fit with a power-law emission superposed to a thermal MEKAL model has
virtually the same reduced $\chi^2$ as the single thermal model fit. In this
fit, the non-thermal flux is only about 5\% of the total flux and the power-law
slope is $\alpha=1.3^{+1.7}_{-1.3}$ ($1\sigma$). This corresponds to the slope
of the relativistic electron energy spectrum of $p=3.6$.

\subsection{Metallicity}

The overall abundance detected with BeppoSAX is very close to the one found by
Pislar et al. (1997) and it is the usual value found in clusters with the same
temperature of Abell~85. The total abundance determined with the MEKAL model is
quite well constrained, mainly by the prominent Fe K$\alpha$ complex at $\sim
6.8$~keV. The mean value obtained for the whole cluster is $0.38 Z_{\odot}$ with
a $3\sigma$ error of 0.06.

The individual iron abundance is also well constrained using the VMEKAL model,
and we systematically obtain a lower value for the Fe abundance compared to the
mean abundance of all metals. The mean value for the whole field is $0.30$ with
a $3\sigma$ error of 0.05. Notice that the Fe abundance and the mean metal
abundance are only marginally compatible with the $3\sigma$ error bar -- these
abundances are different at $2\sigma$ level.

The above result is consistent with the fact that the individual abundances of
other metals (except perhaps for Ni) obtained with the VMEKAL model are usually
higher than the iron abundance (cf. Table~\ref{tbl:abund}). In other words, the
abundance ratio of $\alpha$-elements (Si, S, Ar and Ca) to
iron is greater than one (in solar units). However, the Ni/Fe abundance ratio is
systematically lower than the $\alpha$-elements/Fe. This result must be taken
very cautiously, given the large error bars of the $\alpha$-elements, and is
most significant for the fits of all field and the centre region.

\begin{table}[htb]
\tabcolsep=0.55\tabcolsep
\caption{Abundances relative to iron derived from
Table~\ref{tbl:fitsResults} for all field, the centre region and the first two 
rings. The numbers in brackets give the observed $1\sigma$ interval.}
\begin{tabular}{l|c c c c c}
\cline{2-6}
\multicolumn{1}{l}{} &    Si     &     S     &     Ar    &    Ca     &     Ni\\
\hline
all field &   2.0     &    0.7    &    1.0    &    2.0    &    0.3\\
          & [1.0--3.0]& [0.0--1.7]& [0.0--3.0]& [0.0--4.0]& [0.0--1.7]\\[4pt]

Centre    &   1.8     &   1.5     &   2.5     &   6.0     &    1.0\\
          & [0.8--2.8]& [0.5--2.8]& [0.5--4.8]& [4.0--8.2]& [0.0--2.5]\\[4pt]
	  
1{\small st} Ring  &   2.1     &   ---     &   3.6     &   ---     &    0.4\\
          & [0.7--3.6]&   ---     & [0.4--6.8]&   ---     & [0.0--2.5]\\[4pt]
	  
2{\small nd} Ring  &   4.7     &   2.9     &  ---      &   3.5     &   4.7  \\
          & [1.1--8.2]& [0.0--6.5]&  ---      & [0.0--11.1]& [0.0--10.0]\\
\hline
\end{tabular}
\label{tbl:abund}
\end{table}

The results presented in Table~\ref{tbl:abund} are somewhat in disagreement with
those found by Mushotzky et al. (1996) (see also Loewenstein and Mushotzky 1996)
using ASCA-SIS data for four rich clusters of galaxies (A496, A1060, A2199 and
AWM7). Although we also find an overabundance of Si, our abundance rates for S,
Ar and Ca relative to iron are higher than theirs. The relative abundance of Si
to Fe found here in Abell~85 is in remarkable agreement with the results of
Fukazawa et al. (1998), also based on ASCA data but for 40 clusters. Our
results, Si/Fe~$\approx 2$ and $T = 6.6$~keV, fall squarely on the correlation
between the Si/Fe abundance and cluster temperature (cf. their Fig.~3).

Our results tend to support the burst model for elliptical galaxies, where a
strong galactic wind develops early in the galaxy history (Martinelli et al.
1999). These authors argue that, for the burst model, there should be an
overabundance of $\alpha$-elements compared to iron whereas for the continuous
model, the abundance ratio is smaller than one at $z \sim 0$. The bi-modal model
proposed by Elbaz et al. (1995) also predicts a higher abundance of Si compared
to Fe. Following their conclusion, our results suggest that type II SN have the
main role in the enrichment of the ICM. Fukazawa et al. (1998) also suggest
that, if the Si/Fe relative abundance is indeed correlated with the cluster
temperature, then the role of type II SN should be more important for hotter
clusters.

Based on upper limits on the red spectral line of [Ca \textsc{ii}], Donahue \&
Voit (1993) suggest that Ca is likely depleted onto dust grains in cluster
cooling flows. The higher abundance of calcium derived by us may be the result
of dust evaporation in the hot ICM -- the calcium we observe in X-rays is
completely ionized.

\subsection{Cooling-flow}

We find a lower temperature in the central region, when compared to the mean
cluster temperature, in a similar way that Markevitch et al. (1998) also find.
They claim that it is a sign of a strong central cooling-flow. However, the
difference in temperature we find is only of about $0.5$~keV and, given our
$3\sigma$ error bars, our results are also compatible with an isothermal
temperature radial profile. Nevertheless, at $1\sigma$ level, the temperature is
not isothermal, having a profile that decreases towards the centre. The
outermost ring shows an substantial increase in temperature, but its error bars
are very large (due to the small number of counts).
 
Supporting the existence of a central cooling-flow is the fact that we have a
better fit using a 2-temperature plasma model, with one of the components being
very cold. This cold component was detected only in the central 2~arcmin region,
as observed in most clusters with cooling-flows. However, this cold gas at $T
\la 0.1\,$keV cannot be related to the Extreme Ultra-Violet (EUV) excess found
in some clusters by the EUVE satellite (Lieu et al. 1996; Lieu et al. 1999).

A cooling-flow picture is also supported by the hydrogen column density that we
detect in the centre, much higher that the mean cluster value and at least two
times higher than the galactic value from Dickey \& Lockman (1990). Notice that
there is some correlation between $N_{\rm H}$ and $T$ (as seen in
Fig.~\ref{fig:nH_T_all_centre}) in the sense that lower temperatures are
obtained with higher hydrogen column densities.

The last evidence for a cooling-flow is the large metallicity found in the
centre, significantly higher (at the $3\sigma$ level) than the metallicity found
elsewhere in the cluster.

The central region also shows a strong radio emission, both at 330 and 1400~MHz,
however we detected no non-thermal emission. The superposed radio and
optical data presented in our earlier work (cf. Figure 2, Bagchi et al.~1998)
show that this radio emission comes from the region of the central giant
elliptical and another elliptical $\sim 1.5$ arcmin to north-west of it.
Possibly both these are active radio galaxies pouring relativistic electrons in
the ICM.

\section{Conclusions}

We have presented new X-ray data in the range [0.1--10.0]~keV obtained with
BeppoSAX. Upon analysing the spectrum of a circular region centred at the
position of the VSSRS, we have derived the following mean values:
$T=6.6\pm0.3$~keV, $Z = 0.38\pm0.06Z_{\odot}$ and $N_{\rm H} = 5.5^{+0.9}_{-0.7}
10^{20}$cm$^{-2}$. The temperature is in good agreement with the value
determined by ASCA (Markevich et al. 1998), while the metallicity is in
agreement with the value determined by ROSAT PSPC (Pislar et al. 1997).

Our main results are summarized below:

\begin{itemize}
 \item We have derived an abundance ratio between $\alpha$-elements/iron greater
 than 1 and the iron metallicity is systematically lower than the mean
 metallicity. The over abundance of $\alpha$-elements may be an indication of an
 early enrichment of the ICM by type II SN in elliptical galaxies.

 \item The central region is better fitted with two temperature components, the
 lower temperature being lower than $\sim 0.1$~keV. This supports the existence
 of a multi-phase central cooling-flow.

 \item Since we could not separate unambiguously the thermal from the
 non-thermal X-ray emission, we have estimated, in a number of cases with
 different fitting procedures, the non-thermal X-ray flux. From these values and
 using the radio spectrum data, we derive a lower limit intensity of the
 extended magnetic field, $B \ga 0.9 \mu$G.
    
 \item The South Blob region shows a significantly lower metallicity than the
 rest of the cluster. Its temperature is also higher than its neighbouring
 regions. Such results support the picture of an in-falling stream
 merging with the main body of Abell~85 in that region. The VLA radio
 data shows the presence of both extended magnetic field and relativistic
 particles in this zone. These may originate in an energy transfer from a strong
 shock that formed in a supersonic merger of intergalactic matter.
    
\end{itemize}

\acknowledgements{We thank F. Durret and D. Gerbal for fruitful comments and
discussion. We thank Fabrizio Fiori for his valuable help with the BeppoSAX data
reduction. This research has made use of SAXDAS linearized and cleaned event
files (Rev.1.1) produced at the BeppoSAX Science Data Center. GBLN acknowledges 
financial support from the \textsc{usp/cofecub} cooperation.}

\end{document}